\documentclass[printer]{tCPH2ep}
\usepackage{graphicx}
\usepackage{natbib}
\usepackage{hyperref}

\newcommand{\D}{\ensuremath{\mathcal{D}}}
\def\vev#1{\langle #1\rangle_0}

\newcommand{\ie}{{\em i.e.}}

\newcommand{\gev}{\hbox{ GeV}}

\newcommand{\ev}{\hbox{ eV}}
\newcommand{\mev}{\hbox{ MeV}}

\newcommand{\tev}{\hbox{ TeV}}

\newcommand{\s}{\hbox{ s}}

\newcommand{\fm}{\hbox{ fm}}
\newcommand{\cm}{\hbox{ cm}}

\newcommand{\m}{\hbox{ m}}

\newcommand{\eqn}[1]{(\ref{#1})}
\newcommand{\Eqn}[1]{Eq.~\ref{#1}}
\newcommand{\abs}[1]{\left| #1\right|}

\newcommand{\cfrac}[2]{\textstyle \frac{#1}{#2}}

\def\ltap{\mathop{\raisebox{-.4ex}{\rlap{$\sim$}} 
\raisebox{.4ex}{$<$}}}

\newcommand{\onetev}{1-TeV scale}
\newcommand{\lag}{\ensuremath{\mathcal{L}}}

\newcommand{\ewgg}{\ensuremath{\mathrm{SU(2)_L }\otimes \mathrm{U(1)}_Y}}

\begin{document}

\markboth{Chris Quigg}{Higgs boson \& electroweak symmetry breaking}

\title{Higgs Bosons, Electroweak Symmetry Breaking,\\
 and the Physics of the Large Hadron Collider}

\author{CHRIS QUIGG$^{\ast}$\thanks{$^\ast$Email: quigg@fnal.gov \hfill \textsf{FERMILAB--PUB--07/002--T}
\vspace{6pt}} \\
Theoretical Physics Department, Fermi National Accelerator Laboratory \\
P.O. Box 500, Batavia, Illinois 60510 USA \\ and \\ Theory Group, Physics Department, CERN, CH-1211 Geneva 23, Switzerland
\vspace{6pt}\received{\today} }

\maketitle

\begin{abstract}
The Large Hadron Collider, a $7 \oplus 7\tev$ proton-proton collider under construction at CERN (the European Laboratory for Particle Physics in Geneva), will take experiments squarely into a new energy domain where mysteries of the electroweak interaction will be unveiled. What marks the 1-TeV scale as an important target? Why is understanding how the electroweak symmetry is hidden important to our conception of the world around us? What expectations do we have for the agent that hides the electroweak symmetry? Why do particle physicists anticipate a great harvest of discoveries within reach of the LHC?
\medskip
\noindent \textit{PACS categories:}
{12.15.-y, 14.80.Bn, 11.15.Ex}
 \end{abstract}

\section{Introduction \label{sec:intro}}
Electromagnetism and the weak interactions share a common origin in the weak-isospin and weak-hypercharge symmetries described by the gauge group \ewgg, but their manifestations are very different. Electromagnetism is a force of infinite range, while the influence of the charged-current weak interaction responsible for radioactive beta decay only spans distances shorter than about $10^{-15}\cm$, less than 1\% of the proton radius. We say that the electroweak gauge symmetry is spontaneously broken---hidden---to the $\mathrm{U(1)_{\mathrm{em}}}$ phase symmetry of electromagnetism. How the electroweak gauge symmetry is hidden is one of the most urgent and challenging questions before particle physics.

The search for the agent that hides the electroweak symmetry is also one of the most fascinating episodes in the history of our quest to understand the material world. Over the next decade, experiments at the Large Hadron Collider (LHC) will lead us to a new understanding of  questions that are both simple and profound: Why are there atoms?  Why chemistry?  What makes stable structures possible?  Uncovering the answers to those questions may even bring new insight into ``What makes possible the prerequisites for life?'' A goal of this Article is to link these questions to the electroweak theory, and to the explorations soon to come at the LHC.

Within the standard electroweak theory, the agent of electroweak symmetry breaking is posited to be a single elementary scalar particle known as the Higgs boson, and so ``the search for the Higgs boson'' is a common token for the campaign to understand the origins of electroweak symmetry breaking. Such a shorthand is fine---so long as it is not taken to define a very limited menu of opportunities for discovery.
As we embark upon the LHC adventure, we will need open and prepared minds!

It is often repeated that the discovery of the Higgs boson will reveal the origin of all mass in the Universe. This statement is deeply incorrect---even if we restrict our attention to the luminous matter that is made of familiar stuff.\footnote{We do not know the nature of the dark matter in the Universe, so cannot yet explain how the mass of the dark-matter particles arises.} We can see why the familiar tagline is not right, even before we have reviewed precisely what we mean by  the Higgs boson. 

At each step down the quantum ladder, we understand mass in different terms. In quotidian experience, the mass of an object is the sum of the masses of its parts. At a level we now find so commonplace as to seem trivial, we understand the mass of any atom or molecule in terms of the masses of 
the atomic nuclei, the mass of the electron, and quantum 
electrodynamics.\footnote{In Dirac's 1929 formulation~\cite{PAMD29}, ``The underlying physical laws necessary for the mathematical theory of \ldots the whole of chemistry are thus completely known, and the difficulty is only that the exact application of these laws leads to equations much too complicated to be soluble.''}  And in precise and practical---if not quite 
``first-principle''---terms, we understand the masses of all the 
nuclides in terms of the proton mass, the neutron mass, and our 
knowledge of nuclear forces.

What about the proton and neutron masses? We have learned from Quantum Chromodynamics, the gauge theory of the strong interactions,  that the dominant contribution to the light-hadron masses is not the masses of the quarks of which they are constituted, but the energy stored up in confining the quarks in a tiny volume~\cite{Wilczek:1999be}. Indeed, the masses $m_u$ and $m_d$ of the up and down quarks are only a few MeV. The quark-mass contribution to the 939-MeV mass of an isoscalar nucleon (averaging proton and neutron properties) is only~\cite{Yao:2006px}
\begin{equation}
M_{N}^{q} = 3\,\frac{m_u + m_d}{2} = (7.5\hbox{ to }16.5)\mev\;,
\label{eq:Nmassfromquarks}
\end{equation}
no more than 2\%. Hadrons such as the proton and neutron thus represent \textit{matter of a novel kind.} In contrast to macroscopic matter, atoms, molecules, and nuclei, the mass of a hadron is not equal to the sum of its constituent masses (up to small corrections for binding energy). The quark masses do account for an important detail of our world: The counterintuitive observation that the neutral neutron $(udd)$ is more massive than the charged proton $(uud)$ by $1.29\mev$ is explained by the fact that $m_d$ exceeds $m_u$ by enough to overcome the proton's greater electromagnetic self-energy.

Our most useful tool in the 
strong-coupling regime is QCD formulated on a spacetime lattice.  Calculating the light-hadron 
spectrum from first principles has been one of the main objectives of 
the lattice program, and important strides have been made recently.  
For example, the CP-PACS (Tsukuba) Collaboration's quenched calculation (no dynamical fermions) matches the observed light-hadron spectrum  at the 10\% level~\cite{Aoki:2002fd}. Though small, the discrepancy is larger than the statistical and systematic uncertainties, and so is interpreted as a shortcoming of the quenched approximation. The 
unquenched results now emerging should improve the situation further~\cite{Namekawa:2004bi}, 
and give us new insights into how well---and why!---the simple quark 
model works.

The successful calculation of the hadron spectrum is a remarkable achievement for the theory of quantum chromodynamics and for lattice techniques. In identifying the energy of quark confinement as the origin of the nucleon mass, \textit{we have explained nearly all the visible mass of the Universe,} since the luminous matter is essentially made of protons and neutrons in stars. To excellent approximation, that visible mass of the Universe arises from QCD---not from the Higgs boson.\footnote{The standard model of particle physics (with its generalization to a unified theory of the stong, weak, and electromagnetic interactions) has taught us many fascinating interrelations, including the effect of heavy-quark 
masses on the low-energy value of the strong coupling constant, which sets the scale 
of the light-hadron masses.  For a quick tour, see my \textit{Physics Today} 
article on the top quark~\cite{Quigg:1997uh}; Bob Cahn's \textit{RMP} 
Colloquium~\cite{Cahn:1996ag} takes a more expansive look at connections within the standard 
model.} 

The Higgs boson and the mechanism of electroweak symmetry breaking are nevertheless of capital importance in shaping our world, accounting for the masses of the weak-interaction force particles and---at least in the standard electroweak theory---giving masses to the quarks and leptons. To develop that importance, we shall begin by sketching the electroweak theory and evoking its successes. Then we will address the key question: What would the world be like if there were no Higgs mechanism?

Once having established that the character of electroweak symmetry breaking is a compelling issue, we will consider where the crucial information should be found. The electroweak theory itself points to the energy scale around $1\tev$, or $10^{12}\ev$, and other considerations also single out the \onetev\ as fertile terrain for new physics. Not by coincidence, the Large Hadron Collider will empower experiments to carry out a thorough exploration of the \onetev. We will describe signatures that will be important in the search for the Higgs boson. Then we will argue, independent of any specific mechanism for electroweak symmetry breaking, that (something like) the Higgs boson must exist.
After a brief mention of other new phenomena to be expected on the \onetev, we will close with a short outlook on the decade of discovery ahead.

 \section{Sources of Mass in the Electroweak Theory \label{sec:EWtheory}}
  
\begin{figure}[t!]
\begin{center}
\includegraphics[width=8.0cm]{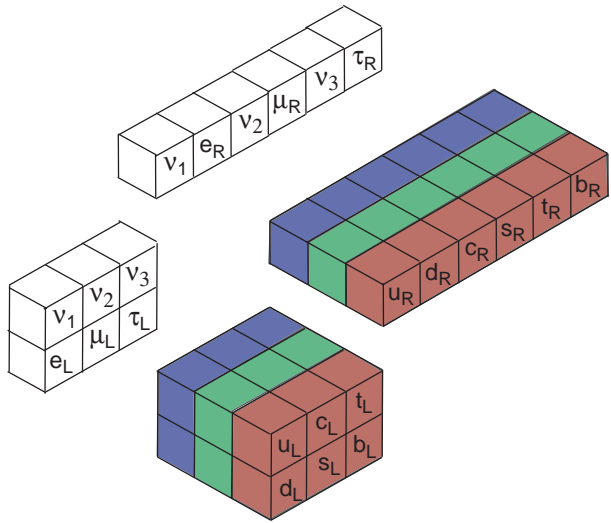}
\caption{Left-handed doublets and right-handed singlets of quarks 
and leptons that inspire 
the structure of the electroweak theory. \label{fig:DumbL}}
\end{center}
\end{figure}

 Our picture of matter is grounded in the recognition of a set of pointlike 
constituents: the quarks and leptons,
as depicted in Figure~\ref{fig:DumbL}, 
plus a few fundamental forces derived from gauge symmetries.\footnote{For general surveys of the standard model of particle physics, and a glimpse beyond, see
\cite{Gaillard:1998ui,CQNewPhys}.} The quarks 
are influenced by the strong interaction, and so carry \textit{color}, 
the strong-interaction charge, whereas the leptons do not feel the 
strong interaction, and are colorless. By pointlike, we understand 
that the quarks and leptons show no evidence of internal structure at 
the current limit of our resolution,  ($r \ltap 10^{-18}\m$). It is striking that the charged-current weak interaction responsible for radioactive beta decay and other processes acts only on the left-handed fermions. We do not know whether the observed parity violation reflects a fundamental asymmetry in the laws of Nature, or a hidden symmetry that might be restored at higher energies. 

Like quantum chromodynamics, the electroweak theory is a gauge theory, in which interactions follow from symmetries. Let us briefly review the strategy of gauge theories by considering the consequences of local gauge invariance in quantum mechanics. A quantum-mechanical state is described by a complex Schr\"{o}dinger wave function $\psi(x)$. Quantum-mechanical observables involve inner products of the form
\begin{equation}
\langle \mathcal{O} \rangle = \int d^n \!x \psi^* \mathcal{O} \psi ,
\label{eqn:observe}
\end{equation}
which are unchanged by a global $\mathrm{U(1)}$ phase rotation, $\psi(x) \to e^{i\theta}\psi(x)$. In other words, the absolute phase of the wave function cannot be measured and is a matter of convention. \textit{Relative} phases between wave functions, as measured in interference experiments, are also unaffected by such a global rotation.

Are we free to choose independent phase conventions in Batavia and in London? In other words, can quantum mechanics be formulated to remain invariant under local, \textit{position-dependent}, phase rotations, $\psi(x) \to e^{i\alpha(x)}\psi(x)$? It is easy to see that this can be achieved, at the price---or reward---of introducing an interaction that we shall construct to be electromagnetism.

Observables such as momentum, as well as  the Sch\"{o}dinger equation itself, involve derivatives of the wave function. Under local phase rotations, these transform as
\begin{equation}
\partial_\mu \psi(x) \to e^{i\alpha(x)}[\partial_\mu \psi(x) + i(\partial_\mu \alpha)\psi(x)] ,
\label{eqn:local}
\end{equation}
which involves more than a mere phase change. The additional gradient-of-phase term, a four-vector, spoils local phase invariance. But we may achieve local phase invariance if we modify the equations of motion and the definitions of observables by introducing the (four-vector) electromagnetic field $A_\mu(x)$. We replace the normal gradient $\partial_\mu$ everywhere---in the Schr\"{o}dinger equation, definition of the momentum operator, etc.---by the \textit{gauge-covariant derivative} $\mathcal{D}_\mu = \partial_\mu +ieA_\mu$, where $e$ is the charge of the particle described by $\psi(x)$. Then if the field $A_\mu(x)$ transforms under local phase rotations as $A_\mu(x) \to A_\mu(x) - (1/e)\partial\alpha(x)$, local phase rotations take
\begin{equation}
\mathcal{D}_\mu \psi(x) \to e^{i\alpha(x)}\mathcal{D}_\mu \psi(x).
\label{eqn:gaugecov}
\end{equation}
Consequently, quantities such as $\psi^{*}\mathcal{D}_\mu \psi$ are invariant under local phase rotations. The transformation law for the four-vector $A_\mu$ has the familiar form of a gauge transformation in electrodynamics. Moreover, the covariant derivative, which prescribes the coupling between matter and the electromagnetic field, corresponds to the replacement $p_\mu \to p_\mu - eA_\mu$ for the momentum operator in the presence of an electromagnetic potential. We have obtained electromagnetism as a consequence of local $\mathrm{U(1)}$ phase invariance applied to the Schr\"{o}dinger wave function.

A parallel strategy can be applied in relativistic quantum field theory for any simple or semi-simple (product) gauge group.
The correct electroweak gauge symmetry emerged through trial and error and experimental guidance. To incorporate electromagnetism into a theory of the weak 
interactions, we add a $\mathrm{U(1)}_{Y}$ weak-hypercharge phase 
symmetry\footnote{We relate the weak hypercharge $Y$ through the 
Gell-Mann--Nishijima connection, $Q = I_{3} + \cfrac{1}{2}Y$, to 
electric charge and (weak) isospin.}
to the $\mathrm{SU(2)_{L}}$ family (weak-isospin) symmetry suggested by 
the left-handed doublets of Figure~\ref{fig:DumbL}. 
To save writing, we shall display the electroweak theory as it 
applies to a single generation of leptons.  In this form, it is incomplete; for quantum corrections to respect the gauge symmetry, a doublet of color-triplet quarks must accompany each doublet of 
color-singlet leptons.  However, the needed generalizations are simple 
enough to make that we need not write them out.\footnote{The electroweak theory is developed in many textbooks; see especially~\cite{Aitchison,CQFIP56,Cough}. For a look back at the evolution of the
 electroweak theory, see the Nobel Lectures by some of its principal
 architects~\cite{Weinberg:1979pi,Salam:1980jd,Glashow:1979pj,'tHooft:2000xn,Veltman:2000xp}.}
We begin by specifying the fermions: a left-handed weak 
isospin doublet
\begin{equation}
{{\sf L}} = \left(\begin{array}{c} \nu_e \\ e \end{array}\right)_{\mathrm{L}}
\end{equation}
with weak hypercharge $Y_{\mathrm{L}}=-1$, and a right-handed weak isospin singlet
\begin{equation}
      {{\sf R}}\equiv e_{\mathrm{R}}
\end{equation}
with weak hypercharge $Y_{\mathrm{R}}=-2$.

The \ewgg\ electroweak gauge group implies two sets of gauge fields:
a weak isovector $\vec{b}_\mu$, with coupling constant $g$, and a
weak isoscalar
${{\mathcal A}}_\mu$, with independent coupling constant $g^\prime$. The gauge fields ensure local gauge invariance, provided they obey the transformation laws $\mathcal{A}_\mu \to \mathcal{A}_\mu - (1/g^\prime)\partial_\mu \alpha$ under an infinitesimal hypercharge phase rotation, and $\vec{b}_\mu \to \vec{b}_\mu - \vec{\alpha} \times \vec{b}_\mu - (1/g)\partial_\mu \vec{\alpha}$ under an infinitesimal weak-isospin rotation generated by $G = 1 + (i/2)\vec{\alpha} \cdot \vec{\tau}$, where $\vec{\tau}$ are the Pauli isospin matrices.
Corresponding
to these gauge fields are the field-strength tensors 
\begin{equation}
    F^{\ell}_{\mu\nu} = \partial_{\nu}b^{\ell}_{\mu} - 
    \partial_{\mu}b^{\ell}_{\nu} + 
    g\varepsilon_{jk\ell}b^{j}_{\mu}b^{k}_{\nu}\; ,
    \label{eq:Fmunu}
\end{equation}
for the weak-isospin symmetry, and 
\begin{equation}
    f_{\mu\nu} = \partial_{\nu}{{\mathcal A}}_\mu - \partial_{\mu}{{\mathcal 
    A}}_\nu \; , 
    \label{eq:fmunu}
\end{equation}
for the weak-hypercharge symmetry.  

We may summarize the interactions 
by the Lagrangian
\begin{equation}
\lag = \lag_{\rm gauge} + \lag_{\rm leptons} \ ,                           
\end{equation}             
with
\begin{equation}
\lag_{\rm gauge}=-\cfrac{1}{4}F_{\mu\nu}^\ell F^{\ell\mu\nu}
-\cfrac{1}{4}f_{\mu\nu}f^{\mu\nu},
\label{eq:gaugeL}
\end{equation}
and
\begin{eqnarray}     
\lag_{\rm leptons} & = & \overline{{\sf R}}\:i\gamma^\mu\!\left(\partial_\mu
+i\frac{g^\prime}{2}{\cal A}_\mu Y\right)\!{\sf R} 
\label{eq:matiere} \\ 
& + & \overline{{\sf
L}}\:i\gamma^\mu\!\left(\partial_\mu 
+i\frac{g^\prime}{2}{\cal
A}_\mu Y+i\frac{g}{2}\vec{\tau}\cdot\vec{b}_\mu\right)\!{\sf L}. \nonumber
\end{eqnarray}
The theory in this form has important shortcomings. 
The Lagrangian of Eq.~\ref{eq:gaugeL}
contains four massless electroweak gauge bosons, namely ${{\mathcal A}}_\mu$, 
$b^{1}_{\mu}$, $b^{2}_{\mu}$, and $b^{3}_{\mu}$, whereas Nature has 
but one: the photon.  (Note that a mass term such as $\cfrac{1}{2}m^2\mathcal{A}_\mu\mathcal{A}^\mu$ is not invariant under a gauge transformation.)
Moreover, the \ewgg\ gauge symmetry forbids a mass term $m\bar{e}e = m(\bar{e}_{\mathrm{R}}e_{\mathrm{L}} + \bar{e}_{\mathrm{L}}e_{\mathrm{R}})$ for the electron in Eq.~\ref{eq:matiere}, because the left-handed and right-handed fields transform differently.  To give masses to the gauge bosons and 
constituent fermions, we must hide the electroweak symmetry, recognizing that a symmetry of the laws of Nature does not imply the same symmetry in the outcomes of those laws.

The most apt analogy for the hiding of the electroweak gauge 
symmetry is found in the superconducting phase transition\footnote{The parallel between electroweak 
symmetry breaking and the Ginzburg-Landau theory is drawn carefully 
in  \S 4.4 of~\cite{Marshak}.  For a rich discussion of 
superconductivity as a consequence of the spontaneous breaking of electromagnetic 
gauge symmetry, see  \S 21.6 of~\cite{SWFT}. For an essay on mass generation through spontaneous symmetry breaking, see~\cite{Wilczek:2000it}.}
To give masses to the intermediate bosons of the weak interaction, we appeal to the Meissner effect---the exclusion of magnetic fields from a superconductor, which corresponds to a nonzero photon mass within the superconducting medium. The Higgs mechanism~\cite{Higgs:1964ia,Englert:1964et,Higgs:1964pj,Higgs:1966ev} is a relativistic generalization of the Ginzburg-Landau 
phenomenology of superconductivity.  We introduce 
a complex doublet of scalar fields
\begin{equation}
\phi\equiv \left(\begin{array}{c} \phi^+ \\ \phi^0 \end{array}\right)
\end{equation}
with weak hypercharge $Y_\phi=+1$.  Next, we add to the Lagrangian new 
(gauge-invariant) terms for the interaction and propagation of the 
scalars,
\begin{equation}
      \lag_{\rm scalar} = (\D^\mu\phi)^\dagger(\D_\mu\phi) - V(\phi^\dagger \phi),
\end{equation}
where the gauge-covariant derivative is
\begin{equation}
      \D_\mu=\partial_\mu 
+i\frac{g^\prime}{2}{\cal A}_\mu
Y+i\frac{g}{2}\vec{\tau}\cdot\vec{b}_\mu \; ,
\label{eq:GcD}
\end{equation}
and (inspired by Ginzburg \& Landau) the potential interaction has the form
\begin{equation}
      V(\phi^\dagger \phi) = \mu^2(\phi^\dagger \phi) +
\abs{\lambda}(\phi^\dagger \phi)^2 .
\label{SSBpot}
\end{equation}
We are also free to add a (gauge-invariant) Yukawa interaction between the scalar fields
and the leptons,
\begin{equation}
      \lag_{\rm Yukawa} = -\zeta_e\left[\overline{{\sf R}}(\phi^\dagger{\sf
L}) + (\overline{{\sf L}}\phi){\sf R}\right].
\label{eq:Yukterm}
\end{equation}

We then arrange 
their self-interactions so that the vacuum state corresponds to a 
broken-symmetry solution.  The electroweak symmetry is spontaneously broken if the parameter
$\mu^2<0$. The minimum energy, or vacuum state, may then be chosen
to correspond to the vacuum expectation value
\begin{equation}
\vev{\phi} = \left(\begin{array}{c} 0 \\ v/\sqrt{2} \end{array}
\right),
\label{eq:vevis}
\end{equation}
where $v = \sqrt{-\mu^2/\abs{\lambda}}$.

Let us verify that the vacuum of Eq.~\ref{eq:vevis} indeed breaks the gauge 
symmetry.  The vacuum state $\vev{\phi}$ is invariant under a symmetry 
operation $\exp{(i \alpha {\mathcal G})}$ corresponding to the 
generator ${\mathcal G}$ provided that $\exp{(i \alpha {\mathcal 
G})}\vev{\phi} = \vev{\phi}$, \ie, if ${\mathcal G}\vev{\phi} = 0$.  
Direct calculation reveals that
 the  original four generators are all broken, but electric charge is
not.  We have accomplished our goal of breaking
$\ewgg \to \mathrm{U(1)}_{\mathrm{em}}$.  The
photon remains massless, but the other three gauge bosons acquire 
masses, as auxiliary scalars assume the role of the third 
(longitudinal) degrees of freedom.  

With the definition $g^{\prime} = g\tan\theta_{W}$, where $\theta_{W}$ 
is the weak mixing angle, we can express the photon as the linear combination
$A = \mathcal{A}\cos{\theta_{W}} + b_{3}\sin{\theta_{W}}$. We identify the strength of its coupling to charged particles, $gg^{\prime}/\sqrt{g^{2} + g^{\prime 2}}$, with the electric charge $e$.
The mediator of the charged-current weak 
interaction, $W^{\pm} = (b_{1} \mp ib_{2})/\sqrt{2}$, acquires a 
mass $M_{W} = gv/2 = ev/2\sin{\theta_{W}}$. The electroweak gauge theory reproduces the low-energy phenomenology of the Fermi theory of weak interactions, provided we set $v = (G_{\mathrm{F}}\sqrt{2})^{-1/2} = 246\gev$, where $G_{\mathrm{F}} = 1.166 37 \times 10^{-5}\gev^{-2}$ is the Fermi constant. It follows at once that $M_W \approx 37.3\gev/\sin{\theta_{W}}$. The combination of $I_3$ and $Y$ orthogonal to the photon is the mediator of the neutral-current weak interaction, $Z = b_{3}\cos{\theta_{W}} - \mathcal{A}\sin{\theta_{W}}$, which
acquires a mass $M_Z=M_W/\cos{\theta_W}$.\footnote{The weak neutral-current interaction was not known before the electroweak theory. Its discovery in 1973~\cite{Haidt:2004ne} marked an important milestone, as did the observation a decade later of the $W^{\pm}$ and $Z^0$ bosons~\cite{Rubbia:1985pv}.}

As a vestige 
of the spontaneous symmetry breaking, there remains a massive, 
spin-zero particle, the Higgs boson.  Its mass  is 
given symbolically as $M_{H}^{2} = -2\mu^{2} > 0$, but \textit{we have no 
prediction for its value.}  Though what we take to be the work of the 
Higgs boson is all around us, the Higgs particle itself has not yet 
been observed!

The masses of the elementary fermions are a more mysterious 
story:
Each fermion mass involves a new, so far incalculable, 
Yukawa coupling.  
When the electroweak symmetry is spontaneously broken, the electron 
mass emerges as
$m_{e} = \zeta_{e}v/\sqrt{2}$. The Yukawa couplings that reproduce the observed quark and lepton 
masses range over many orders of magnitude, from $\zeta_{e} \approx 
3 \times 10^{-6}$ for the electron to $\zeta_{t} \approx 1$ for the 
top quark.  Their origin is unknown.
In that sense, therefore, \textit{all fermion masses involve physics
beyond the standard model.}

\begin{figure}[tb]
	\centerline{\includegraphics[width=8.0cm]{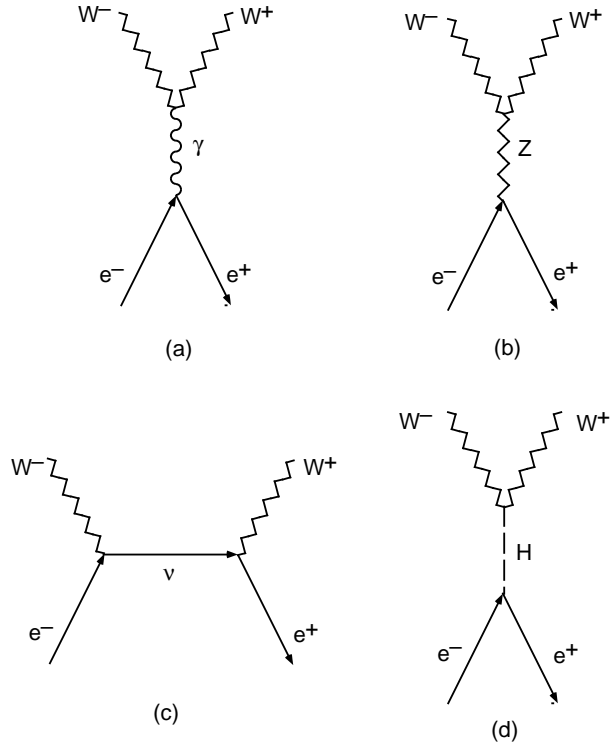}}
	\vspace*{6pt}
	\caption{Lowest-order contributions to the $e^+e^- \rightarrow 
	W^{+}W^{-}$ scattering amplitude.}
	\protect\label{fig:eeWW}
\end{figure}

\begin{figure}[tb]
	\centerline{\includegraphics[width=8.0cm]{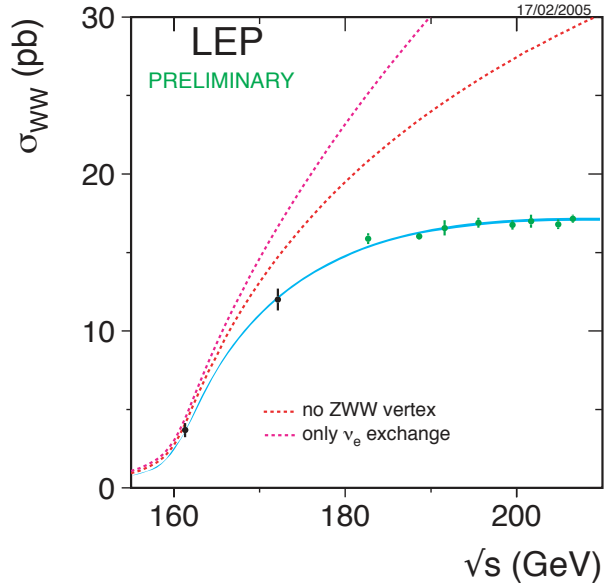}}
	\vspace*{6pt}
	\caption{Cross section for the reaction $e^{+}e^{-} \to W^{+}W^{-}$ 
	measured by the four LEP experiments, together with the full 
	electroweak-theory simulation and the cross sections that would 
	result from $\nu$-exchange alone and from $(\nu+\gamma)$-exchange
	{\protect \cite{lepewwg}}.}
	\protect\label{fig:LEPgc}
\end{figure}

Experiments and the supporting theoretical calculations over the 
 past decade have elevated the electroweak theory to a law of Nature, tested as a quantum field theory at the level of one per mille.\footnote{The current state of the theory is reviewed in~\cite{deJong:2005mk}.
An ongoing comparison of theory and experiment is maintained 
 by the LEP Electroweak Working Group~\cite{lepewwg}.} One
  remarkable achievement of recent experiments is a clear 
test of the gauge symmetry, or group-theory structure, of the 
electroweak theory, in the reaction $e^{+}e^{-} \to W^{+}W^{-}$. 
Neglecting the electron mass, this reaction is described by three 
Feynman diagrams that correspond to  $s$-channel photon and $Z^{0}$ exchange, and $t$-channel neutrino exchange, Figure \ref{fig:eeWW}(a-c). 
Each diagram leads to a $J = 1$ partial-wave amplitude $\propto s$, the square of the c.m.\ energy, but the gauge symmetry enforces a pattern of cooperation:
 The contributions of the direct-channel 
$\gamma$- and $Z^0$-exchange diagrams 
of Figs.~\ref{fig:eeWW}(a) and (b) cancel the leading divergence in the $J=1$ 
partial-wave amplitude of 
the neutrino-exchange diagram in Figure~\ref{fig:eeWW}(c).  The LEP measurements~\cite{lepewwg} plotted in Figure~\ref{fig:LEPgc}
agree well with the benign high-energy behavior predicted by the electroweak theory. If the 
$Z$-exchange contribution is omitted (middle line) or if both the 
$\gamma$- and $Z$-exchange contributions are omitted (upper
line), the calculated cross section grows unacceptably with 
energy---and disagrees with the measurements.  The gauge cancellation 
in the $J=1$ partial-wave amplitude is thus observed.

 \section{Why Electroweak Symmetry Breaking Matters to You \label{sec:itmatters}}
Experiments that explore the \onetev\ will deepen our understanding of the everyday, the stuff of the world around us, responding in new and revealing ways to the basic questions about atoms, chemistry, and complex objects announced in \S\ref{sec:intro}. Perhaps the best way to connect those questions with the electroweak theory and LHC physics  is to consider what the world would be
like if there were nothing like the Higgs mechanism for electroweak
symmetry breaking.  
First, it's clear that quarks and leptons would
remain massless, because mass terms are not permitted  if the electroweak symmetry remains
manifest.\footnote{I assume for this discussion that all the trappings
of the Higgs mechanism, including Yukawa couplings for the fermions,
are absent.} Eliminating the Higgs mechanism does nothing to the strong interaction, so QCD would
still confine the (massless) color-triplet quarks into color-singlet
hadrons, with very little change in the masses of those stable
structures.  

An interesting and slightly subtle point is that, even in the
absence of a Higgs mechanism, QCD hides the electroweak symmetry~\cite{Weinstein:1973gj}.  
In a world with massless up and down quarks, QCD exhibits a global $\mathrm{SU(2)_L\otimes SU(2)_R}$ \textit{chiral symmetry} that treats the left-handed and right-handed quarks
as separate objects. As we approach low energy from above, that chiral
symmetry  is spontaneously broken.  The resulting communication between the
left-handed and right-handed worlds engenders a breaking of the
electroweak symmetry: $\mathrm{SU}(2)_{\mathrm{L}}\otimes \mathrm{U}(1)_{Y}$
becomes $\mathrm{U}(1)_{\mathrm{em}}$, and the gauge bosons are the massless photon and massive $W^\pm$ and $Z^0$. Despite the structural similarity to the standard model, this is not a satisfactory theory of the weak interactions. Here the scale of electroweak symmetry breaking is measured by the pion lifetime---the coupling of the axial current to the vacuum. The amount of mass acquired
by the $W$ and $Z$ is too small by a factor of 2500.

Because the weak bosons have acquired
masses, the weak-isospin force, which we might have taken to be a
confining force in the absence of symmetry breaking, does not confine objects bearing weak isospin into weak-isospin singlets. The familiar spectrum of hadrons persists, but with a crucial difference. The proton, with its electrostatic self-energy, will now outweigh the neutron, because in this world of massless quarks, the down quark now does not outweigh the up quark.

Beta decay---exemplified in this world by $p \to n e^+ \nu_e$---is very rapid, because the gauge bosons are so light.  The
lightest nucleus is therefore one neutron; \textit{there is no hydrogen atom.} 
Exploratory analyses of what would happen to big-bang
nucleosynthesis in this world suggest that some light elements, such as helium, would be created in the first minutes after the big bang ~\cite{PhysRevLett.80.1822,PhysRevD.57.5480,RevModPhys.72.1149,PhysRevD.67.043517}. [It would be interesting to see this worked out in complete detail.]
Because the electron is
massless, the Bohr radius of the atom is infinite, so there is nothing
we would recognize as an atom, there is no chemistry as we know it,
there are no stable composite structures like the solids and liquids we
know.\footnote{It is nearly inevitable that effects negligible in our world would, in the Higgsless world, produce fermion masses. These are typically many orders of magnitude smaller than the observed masses, small enough that the Bohr radius of a would-be atom would be macroscopic, sustaining the conclusion that matter would lose its integrity.}

How very different the world would be, were it not for the
mechanism of electroweak symmetry breaking!  What we are really trying
to get at, when we look for the source of electroweak symmetry
breaking, is why we don't live in a world so different, why we live in
the world we do.  This is one of the
deepest questions that human beings have ever tried to engage, and
it is coming within the reach of particle physics.

What form might the answer take?  What clues we have suggest that the agent of electroweak symmetry breaking represents a novel fundamental interaction at an energy of a few
hundred GeV. \textit{We do not know what that force is.} 

 It could be the Higgs
mechanism of the standard model (or a supersymmetric elaboration of the standard model~\cite{Martin:1997ns}),
which is built in analogy to the Ginzburg--Landau description of
superconductivity.  The potential that we arrange, by decree, to hide the electroweak symmetry arises not from gauge forces but from an entirely new kind of interaction.

Maybe the electroweak symmetry is hidden by a new gauge force. One very 
appealing possibility---at least until you get into the details---is 
that the solution to electroweak symmetry breaking will be like the 
solution to the model for electroweak symmetry breaking, the 
superconducting phase transition. The superconducting phase transition 
is first described by the Ginzburg--Landau phenomenology, but then in 
reality is explained by the Bardeen--Cooper--Schrieffer theory that 
comes from the gauge theory of Quantum Electrodynamics. Maybe, then, 
we will discover a mechanism for electroweak symmetry breaking almost 
as economical as the QCD mechanism we discussed above. One much investigated line is the possibility that 
new constituents still to be discovered interact by 
means of yet unknown forces, and when we learn how to 
calculate the consequences of that theory we will find our analogue 
of the BCS theory~\cite{Hill:2002ap}.  

It could even be that there is some truly
emergent description of the electroweak phase
transition, a residual force that arises from the strong dynamics among
the weak gauge bosons~\cite{Chanowitz:2004gk}.  If we take the mass of the Higgs
boson to very large values (beyond $1\tev$ in the Lagrangian of the
electroweak theory), the scattering among gauge bosons becomes strong,
in the sense that $\pi\pi$ scattering becomes strong on the GeV scale, as we shall see in \S\ref{sec:import}. In that event, it is reasonable to speculate that resonances form among pairs of gauge bosons, multiple production of
gauge bosons becomes commonplace, and that resonant behavior could hold the key to understanding what hides the electroweak symmetry.  

Much model building has occurred around the proposition that  the Higgs boson is a pseudo-Nambu-Goldstone boson of a spontaneously broken approximate global symmetry, with the explicit breaking of this symmetry collective in nature, that is, more than one coupling at a time must be turned on for 
the symmetry to be broken. These ``Little Higgs'' theories feature weakly coupled new physics at the TeV scale~\cite{Schmaltz:2005ky}.

Or perhaps electroweak symmetry breaking is an echo of extra spacetime dimensions. Among the possibilities are models without a physical Higgs scalar, in which is electroweak symmetry is hidden by boundary conditions~\cite{Csaki:2005vy}.

Theory has offered many alternatives. During the next decade, experiment will tell us which path
Nature has taken. An essential first step is to find the Higgs boson and to learn its properties. But where shall we look?

\section{Higgs-Boson Properties}
Once we assume a value for the Higgs-boson mass, it is a simple matter 
to compute the rates for Higgs-boson decay into pairs of fermions or 
weak bosons.  For a fermion $f$ with $N_{c}$ colors, the partial width $\Gamma(H \to f\bar{f})$ is
 proportional to $N_c m_f^2 M_{H}$ in the limit of large Higgs mass.
 The rates for decays into 
weak-boson pairs are asymptotically proportional to $M_{H}^{3}$ and 
$\cfrac{1}{2}M_{H}^{3}$, for $H \to W^+W^-$ and $H \to ZZ$, respectively.  The 
dominant decays for large $M_{H}$ are into pairs of longitudinally 
polarized weak bosons.

\begin{figure}[tb]
	\centerline{\includegraphics[width=10.0cm]{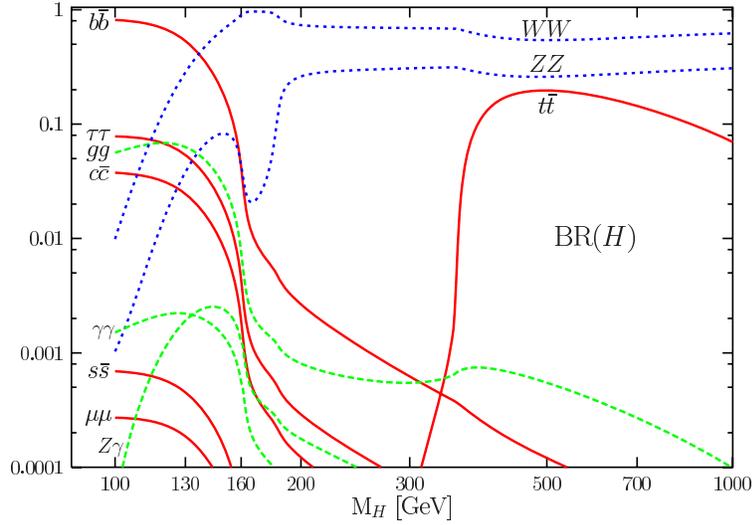}}
	\vspace*{6pt}
	\caption{Branching fractions for the prominent decay modes of the standard-model 
	Higgs boson, from~\cite{Djouadi:2005gi}.}
	\protect\label{fig:LHdk}
\end{figure}

Branching fractions for decay modes that may hold promise for the 
detection of a Higgs boson are displayed in Figure 
\ref{fig:LHdk}.  In addition to the $f\bar{f}$ and $VV$ modes that 
arise at tree level, the plot includes the $\gamma\gamma$, $Z\gamma$, and two-gluon modes that 
proceed through loop diagrams.  Though rare, the $\gamma\gamma$ 
channel offers an important target for LHC experiments, if the Higgs boson is light.

\begin{figure}[t!]
	\centerline{\includegraphics[width=10.0cm]{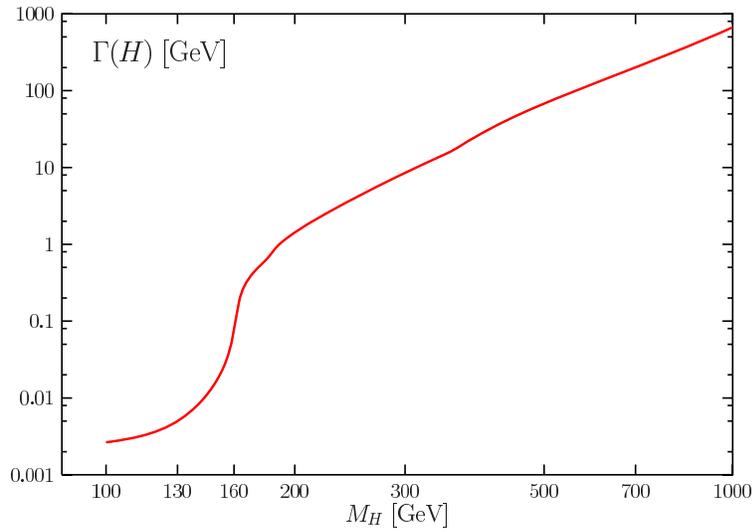}}
	\vspace*{6pt}
	\caption{Higgs-boson total width as a function of mass, from~\cite{Djouadi:2005gi}.}
	\protect\label{fig:Htot}
\end{figure}

Below the $W^{+}W^{-}$ threshold, the total width of the 
standard-model Higgs boson is rather small, typically less than 
$1\gev$.  Far above the threshold for decay into gauge-boson pairs, 
the total width is proportional to $M_{H}^{3}$.  At masses 
approaching $1\tev$, the Higgs boson becomes very broad, with a 
perturbative width approaching its mass.  The Higgs-boson total width 
is plotted as a function of $M_{H}$ in Figure \ref{fig:Htot}.

As we have described it, the Higgs boson is an artifact of the mechanism we chose to hide the electroweak symmetry. What assurance do we have that a Higgs boson, or something very like it, will be 
found? One path to the \emph{theoretical} discovery of the Higgs boson
involves its role in the cancellation of 
high-energy divergences. We saw at the end of \S\ref{sec:EWtheory} that the most severe divergences of the individual $\nu$-, $\gamma$-, and $Z$-exchange diagrams in the reaction $e^+ e^- \to W^+ W^-$ are tamed by a cooperation among the three diagrams of Figure~\ref{fig:eeWW}(a-c) that follows from gauge symmetry.
However, this is not the end of the high-energy story: the $J=0$
partial-wave amplitude, which exists in this case because the
electrons are massive and may therefore be found in the ``wrong''
helicity state, grows as the c.m.\ energy for the production of
longitudinally polarized gauge bosons.  This unacceptable high-energy behavior is
precisely cancelled by the Higgs boson graph of
Figure~\ref{fig:eeWW}(d).  If the Higgs boson did not exist, something
else would have to play this role.  From the point of view of
$S$-matrix analysis, the Higgs-electron-electron coupling must be
proportional to the electron mass, because the strength of ``wrong-helicity''
configurations is measured by the fermion mass.

Let us underline this result.
If the gauge symmetry were unbroken, there would be 
no Higgs boson, no longitudinal gauge bosons, and no extreme divergence 
difficulties. But there would be no viable low-energy phenomenology
 of the 
weak interactions. The most severe divergences of individual diagrams 
are eliminated by the gauge 
structure of the couplings among gauge bosons and leptons. A lesser, but 
still potentially fatal, divergence arises because the electron has 
acquired mass---because of the Higgs mechanism. Spontaneous symmetry 
breaking provides its own cure by supplying a Higgs boson to remove the 
last divergence. A similar interplay and compensation must exist in any 
satisfactory theory. There will be (almost surely) a spin-zero object
that has effectively more or less the interactions of the
standard-model Higgs boson, whether it be an elementary particle that
we build into to the theory or something that emerges from the theory.
Such an object is required to make the electroweak theory behave well
at high energies, once electroweak symmetry is hidden. 

It is by no means guaranteed that the same agent hides electroweak symmetry and generates fermion mass. We saw in \S\ref{sec:itmatters} that chiral symmetry breaking in QCD could hide the electroweak symmetry without generating fermion masses. In extended technicolor models~\cite{Dimopoulos:1979es,Eichten:1979ah}, for example, separate gauge interactions hide the electroweak symmetry and communicate the broken symmetry to the quarks and leptons. In supersymmetric models, five Higgs bosons are expected, and the branching fractions of the lightest one may be very different from those presented in Figure~\ref{fig:LHdk}~\cite{Djouadi:2005gj}. Accordingly, it will be of great interest to map the decay pattern of the Higgs boson, once it is found, in order to characterize the mechanism of electroweak symmetry breaking.

 \section{The Importance of the 1-TeV Scale \label{sec:import}}
The electroweak theory does not give a precise prediction for the mass of the Higgs boson, but a unitarity argument~\cite{Lee:1977eg} leads to a conditional upper bound on the Higgs 
boson mass that sets a key target for experiment. 

It is straightforward to compute the 
amplitudes ${\cal M}$ for gauge boson scattering at high energies, and to make
a partial-wave decomposition, according to ${\cal M}(s,t)=16\pi\sum_J(2J+1)a_J(s)P_J(\cos{\theta})$.
 Most channels ``decouple,'' in the sense 
that partial-wave amplitudes are small at all energies (except very
near the particle poles, or at exponentially large energies), for
any value of the Higgs boson mass $M_H$. Four channels are interesting:
\begin{equation}
\begin{array}{cccc}
W_L^+W_L^-\quad & {\displaystyle \frac{Z_L^0Z_L^0}{\sqrt{2}}}\quad & {\displaystyle \frac{HH}{\sqrt{2}}}\quad & HZ_L^0 \; ,
\end{array}
\end{equation}
where the subscript $L$ denotes the longitudinal polarization
states, and the factors of $\sqrt{2}$ account for identical particle
statistics. For these, the $s$-wave amplitudes are all asymptotically
constant (\ie, well-behaved) and  
proportional to $G_{\mathrm{F}}M_H^2.$ In the high-energy 
limit,\footnote{It is convenient to calculate these amplitudes by 
means of the Goldstone-boson equivalence theorem~\cite{Cornwall:1974km}, which 
reduces the dynamics of longitudinally polarized gauge bosons to a 
scalar field theory with interaction Lagrangian given by 
$\mathcal{L}_{\mathrm{int}} = -\lambda v h 
(2w^{+}w^{-}+z^{2}+h^{2}) - 
(\lambda/4)(2w^{+}w^{-}+z^{2}+h^{2})^{2}$, with $1/v^{2} = 
G_{\mathrm{F}}\sqrt{2}$ and $\lambda = G_{\mathrm{F}}M_{H}^{2}/\sqrt{2}$.}
\begin{equation}
\lim_{s\gg M_H^2}(a_0)\to\frac{-G_{\mathrm{F}} M_H^2}{4\pi\sqrt{2}}\cdot \left[
\begin{array}{cccc} 1 & 1/\sqrt{8} & 1/\sqrt{8} & 0 \\
      1/\sqrt{8} & 3/4 & 1/4 & 0 \\
      1/\sqrt{8} & 1/4 & 3/4 & 0 \\
      0 & 0 & 0 & 1/2 \end{array} \right] \; .
\end{equation} 
Requiring that the largest eigenvalue respect the 
partial-wave unitarity condition $\abs{a_0}\le 1$ yields
\begin{equation}
	M_H \le \left(\frac{8\pi\sqrt{2}}{3G_{\mathrm{F}}}\right)^{1/2} =1\tev
\end{equation}
as a condition for perturbative unitarity.

If the bound is respected, weak interactions remain weak at all
energies, and perturbation theory is everywhere reliable. If the
bound is violated, perturbation theory breaks down, and weak
interactions among $W^\pm$, $Z$, and $H$ become strong on the \onetev.
This means that the features of strong interactions at GeV energies
will come to characterize electroweak gauge boson interactions at
TeV energies. We interpret this to mean that new phenomena are to
be found in the electroweak interactions at energies not much larger
than 1~TeV.

If the \ewgg\ electroweak theory points to a Higgs boson mass below $1\tev$, it does not explain how the 
scale of electroweak symmetry breaking is maintained in the presence 
of quantum corrections.   Beyond the classical approximation, scalar mass parameters receive 
quantum corrections from loops that contain particles of spins 
$J=0, \cfrac{1}{2}$, and $1$:
\begin{displaymath}
\includegraphics[width=10.0cm]{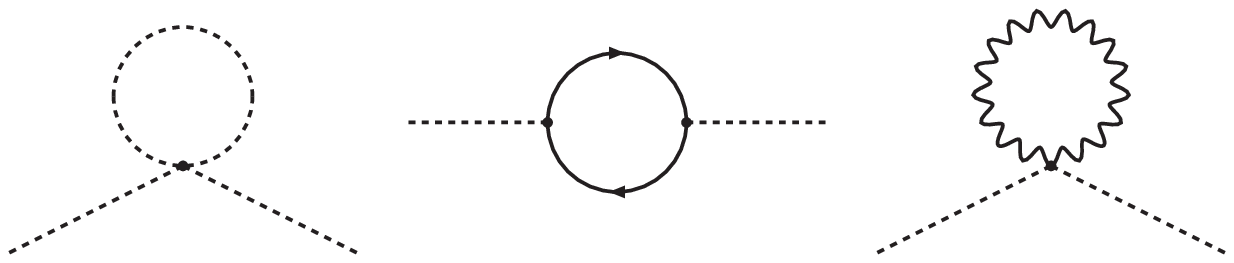}
\label{loup}
\end{displaymath}
The loop integrals are potentially divergent.  Symbolically, we may 
summarize their implications as
\begin{equation}
	M_H^2(p^2) = M_H^2(\Lambda^2) + Cg^2\int^{\Lambda^2}_{p^2}dk^2 
	+ \cdots \;,
	\label{longint}
\end{equation}
where $\Lambda$ defines a reference scale at which the value of 
$M_H^2$ is known, $g$ is the coupling constant of the theory, and the 
coefficient $C$ is calculable in any particular theory.  
Here we describe the variation of an observable with the momentum scale. The loop integrals appear to be quadratically divergent, $\propto \Lambda^2$. 
In order for the mass shifts 
induced by quantum corrections to remain under control (\ie , not to 
greatly exceed the value measured on the laboratory scale), either 
$\Lambda$ must be small, so the range of integration is not 
enormous, or new physics must intervene to damp the integrand.

If the fundamental interactions are described by an 
$\mathrm{SU(3)_c}\otimes \ewgg$ gauge symmetry, \ie, by quantum
chromodynamics and the electroweak theory, then a 
natural reference scale is the Planck mass,
$\Lambda \sim M_{\rm Planck}  = 
	\left({\hbar c}/{G_{\mathrm{Newton}}}\right)^{1/2} \approx 1.2 
	\times 10^{19}\gev$.
In a unified theory of the strong, weak, and electromagnetic 
interactions, a natural scale is the unification scale,
$\Lambda \sim M_U \approx 10^{15}\hbox{ - }10^{16}\gev$.
 Both estimates are very large compared to the electroweak scale.  The challenge of preserving widely separated scales in the presence of quantum corrections is known as the \textit{hierarchy problem.} Unless we suppose that $M_H^2(\Lambda^2)$ and the quantum corrections are finely tuned to yield $M_H^2(p^2) \ltap (1\tev)^2$, some new physics must intervene at an energy of approximately $1\tev$ to bring the integral in \Eqn{longint} under control. 

Note the implications: The unitarity argument showed that new physics must be present on the \onetev, either in the form of a Higgs boson, or other new phenomena. But a low-mass Higgs boson is imperiled by quantum corrections. New physics not far above the \onetev\ could bring the reference scale $\Lambda$ low enough to mitigate the threat. If the reference scale is indeed very large, then either various contributions to the Higgs-boson mass must be precariously balanced or new physics must control the contribution of the integral in \Eqn{longint}. We do not have a proof that Nature is not fine tuned, but I think it highly likely that \textit{both a Higgs boson and other new phenomena} are to be found near the \onetev.

Let us review some of the ways in which new phenomena could resolve the hierarchy problem.
  Exploiting the fact 
that fermion loops contribute with an overall minus sign relative to boson loops (because of 
Fermi statistics), \textit{supersymmetry} balances the contributions of fermion 
and boson loops.\footnote{``Little Higgs" models and ``twin Higgs'' models~\cite{Chacko:2005pe} employ  different conspiracies of contributions to defer the hierarchy problem to about $10\tev$.}  In the limit of unbroken supersymmetry, in which the 
masses of bosons are degenerate with those of their fermion 
counterparts, the cancellation is exact.
If the supersymmetry is broken (as it must be in our world), the 
contribution of the integrals may still be acceptably small if the 
fermion-boson mass splittings $\Delta M$ are not too large.  The 
condition that $g^2\Delta M^2$ be ``small enough'' leads to the 
requirement that superpartner masses be less than about 
$1\tev$. It is provocative to note that, with superpartners at $\mathcal{O}(1\tev)$, the $\mathrm{SU(3)_c}\otimes \ewgg$ coupling constants run to a common value at a unification scale of about $10^{16}\gev$~ \cite{Amaldi:1991cn}.   Theories of dynamical symmetry breaking, such as Technicolor, offer a  second solution to the problem of the enormous range of integration in 
\eqn{longint}. In technicolor models, the Higgs boson is composite, and its internal structure comes into play  on the scale of its binding, $\Lambda_{\mathrm{TC}} \simeq 
\mathcal{O}(1~{\rm TeV})$. The integrand is damped, the effective range of integration is cut off, and 
mass shifts are under control.

We have one more independent indication that new phenomena should be present on the \onetev. An appealing interpretation of the evidence that dark matter makes up roughly one-quarter of the energy density of the Universe~\cite{Spergel:2006hy} is that dark matter consists of  thermal relics of the big bang, stable---or exceedingly long-lived---neutral particles. If the particle has couplings of weak-interaction strength, then generically the observed dark-matter density results if the mass of the dark-matter particle lies between approximately $100\gev$ and $1\tev$~\cite{Bertone:2004pz}. Typically, scenarios to extend the electroweak theory and resolve the hierarchy problem---whether based on extra dimensions, new strong dynamics, or supersymmetry---entail dark-matter candidates on the \onetev. One aspect of the great optimism with which particle physicists contemplate the explorations under way at Fermilab's Tevatron and soon to be greatly extended at CERN's Large Hadron Collider is a strong suspicion that many of the outstanding problems of particle physics and cosmology may be linked---and linked to the \onetev. Dark matter is a perfect example.
  
 \section{Outlook}
 
 Over the next decade, experiments will carry out definitive explorations of the Fermi scale, at energies around $1\tev$ for collisions among quarks and leptons. This is physics on the \textit{nanonanoscale,} probing distances smaller than $10^{-18}\m$. In this regime, we confidently expect to find the key to the mechanism that drives electroweak symmetry breaking, with profound implications for our conception of the everyday world. A pivotal step will be the search for the Higgs boson and the elaboration of its properties. What is more, the hierarchy problem leads us to suspect that other new phenomena are to be found on the \onetev, phenomena that will give new insight into why the electroweak scale is so much smaller than the Planck scale. We also have reason to believe---from arguments about relic densities and also from specific models---that a weakly interacting class of dark-matter candidates could populate the same energy range.
 
 \textit{We do not know what the new wave of exploration will find,} but the discoveries and new puzzles are certain to change the face of particle physics and reverberate through neighboring disciplines. Resolving the conundrums of the \onetev\ should aid us in reformulating some of today's fuzzy questions and give us a clearer view of the physics at still shorter distances, where we may uncover new challenges to our understanding. We could well find new clues to the unification of forces or indications for a rational pattern of constituent masses, viewed at a high energy scale. I hope that we will be able to sharpen the problem of identity---what makes an electron an electron, a top quark a top quark, a neutrino a neutrino---so we can formulate a strategy to resolve it.
 
 Experiments at the Fermilab Tevatron, a 2-TeV proton-antiproton collider, have begun to approach the \onetev. The CDF~\cite{cdf} and D\O~\cite{d0} experiments, which discovered the 171-GeV top quark in 1995, are profiting from world-record machine performance: initial luminosities exceeding $2.5 \times 10^{32}\cm^2\s^{-1}$ and an integrated luminosity to date of more than $2\fm^{-1}$~\cite{tevlum}. They are expected to run through 2009, continuing their pursuit of the Higgs boson, supersymmetry, and other new phenomena.
 
 The Large Hadron Collider~\cite{lhc}, a 14-TeV proton-proton collider at CERN, will transport us into the heart of the Fermi scale. The LHC should demonstrate 900-GeV collisions by the end of 2007 and produce the first 14-TeV events during 2008.\footnote{For a prospectus on early running at the LHC, see~\cite{Gianotti:2005fm}.} Its collision rate will grow to 100 times the Tevatron's luminosity. Like the machine itself, with its 27-km circumference, the multipurpose detectors  
 ATLAS~\cite{atlas} and  CMS~\cite{cms} are both titans and engineering marvels. They will be the vessels for a remarkable era of exploration and discovery.

\section*{Acknowledgments \label{ackno}}
Fermilab is operated by Fermi Research Alliance, LLC  under Contract No.~DE-AC02-07CH11359 with the United States Department of Energy.  It is a pleasure to thank Luis \'{A}lvarez-Gaum\'{e} and other members of the CERN Theory Group for warm hospitality in Geneva. I am grateful to Frank Close, Bogdan Dobrescu, and Joe Lykken for thoughtful comments on the manuscript.

\bibliography{CPHiggs}

\begin{thebibliography}{0}
\providecommand{\natexlab}[1]{#1}

\bibitem{PAMD29}
 P.~A.~M. Dirac, Proc. Royal Soc. (London) \textbf{A123} 714--733 (1929).

\bibitem{Wilczek:1999be}
 F. Wilczek, Phys. Today \textbf{52N11} 11--13 ({November 1999}).

\bibitem{Yao:2006px}
 W.~M. Yao {\itshape et~al.}, J. Phys. \textbf{G33} 1--1232 (2006).

\bibitem{Aoki:2002fd}
 S. Aoki {\itshape et~al.}, Phys. Rev. \textbf{D67} 034503 (2003)
  [\href{http://arXiv.org/abs/hep-lat/0206009}{{\tt hep-lat/0206009}}].

\bibitem{Namekawa:2004bi}
 Y. Namekawa {\itshape et~al.}, Phys. Rev. \textbf{D70} 074503 (2004)
  [\href{http://arXiv.org/abs/hep-lat/0404014}{{\tt hep-lat/0404014}}].

\bibitem{Quigg:1997uh}
 C. Quigg, Phys. Today \textbf{50N5} 20--26 ({May 1997})
  [\href{http://arXiv.org/abs/hep-ph/9704332}{{\tt hep-ph/9704332}}].

\bibitem{Cahn:1996ag}
 R.~N. Cahn, Rev. Mod. Phys. \textbf{68} 951--960 (1996).

\bibitem{Gaillard:1998ui}
 M.~K. Gaillard,  P.~D. Grannis and  F.~J. Sciulli, Rev. Mod. Phys. \textbf{71}
  S96--S111 (1999) [\href{http://arXiv.org/abs/hep-ph/9812285}{{\tt
  hep-ph/9812285}}].

\bibitem{CQNewPhys}
 C. Quigg, Particles and the Standard Model, in G.~Fraser (Editor)  {\itshape
  The New Physics: for the 21st century}  (Cambridge University Press,
  Cambridge \& New York, 2006), chap.~4, pp. 86--118.

\bibitem{Aitchison}
 I.~J.~R. Aitchison and  A.~J.~G. Hey, {\itshape Gauge Theories in Particle
  Physics}, third edition ,  Vol. 2   (Taylor \& Francis, London, 2003).

\bibitem{CQFIP56}
 C. Quigg, {\itshape Gauge Theories of the Strong, Weak, and Electromagnetic
  Interactions}  (Westview Press, Boulder, Colorado, 1997).

\bibitem{Cough}
 G.~D. Coughlan,  J.~E. Dodd and  B.~M. Gripaios, {\itshape The Ideas of
  Particle Physics}  (Cambridge University Press, Cambridge \& New York, 2006).

\bibitem{Weinberg:1979pi}
 S. Weinberg, Rev. Mod. Phys. \textbf{52} 515--523 (1980).

\bibitem{Salam:1980jd}
 A. Salam, Rev. Mod. Phys. \textbf{52} 525--538 (1980).

\bibitem{Glashow:1979pj}
 S.~L. Glashow, Rev. Mod. Phys. \textbf{52} 539--543 (1980).

\bibitem{'tHooft:2000xn}
 G. {'t Hooft}, Rev. Mod. Phys. \textbf{72} 333--339 (2000).

\bibitem{Veltman:2000xp}
 M.~J.~G. Veltman, Rev. Mod. Phys. \textbf{72} 341--349 (2000).

\bibitem{Marshak}
 R.~E. Marshak, {\itshape Conceptual Foundations of Modern Particle Physics}
  (World Scientific, Singapore, 1993).

\bibitem{SWFT}
 S. Weinberg, {\itshape The Quantum Theory of Fields},  Vol. 2   (Cambridge
  University Press, Cambridge, 1996).

\bibitem{Wilczek:2000it}
 F. Wilczek, Phys. Today \textbf{53N1} 13--14 ({January 2000}).

\bibitem{Higgs:1964ia}
 P.~W. Higgs, Phys. Lett. \textbf{12} 132--133 (1964).

\bibitem{Englert:1964et}
 F. Englert and  R. Brout, Phys. Rev. Lett. \textbf{13} 321--322 (1964).

\bibitem{Higgs:1964pj}
 P.~W. Higgs, Phys. Rev. Lett. \textbf{13} 508--509 (1964).

\bibitem{Higgs:1966ev}
 P.~W. Higgs, Phys. Rev. \textbf{145} 1156--1163 (1966).

\bibitem{Haidt:2004ne}
 D. Haidt, Eur. Phys. J. \textbf{C34} 25--31 (2004).

\bibitem{Rubbia:1985pv}
 C. Rubbia, Rev. Mod. Phys. \textbf{57} 699--722 (1985).

\bibitem{deJong:2005mk}
 S. {de Jong}, PoS \textbf{HEP2005} 397 (2006)
  [\href{http://arXiv.org/abs/hep-ex/0512043}{{\tt hep-ex/0512043}}].

\bibitem{lepewwg}
{The LEP Electroweak Working Group}
  [\href{http://lepewwg.web.cern.ch}{\texttt{lepewwg.web.cern.ch}}].

\bibitem{Weinstein:1973gj}
 M. Weinstein, Phys. Rev. \textbf{D8} 2511 (1973).

\bibitem{PhysRevLett.80.1822}
 V. Agrawal,  S.~M. Barr,  J.~F. Donoghue, {\itshape et~al.}, Phys. Rev. Lett.
  \textbf{80} 1822--1825 (1998).

\bibitem{PhysRevD.57.5480}
 V. Agrawal,  S.~M. Barr,  J.~F. Donoghue, {\itshape et~al.}, Phys. Rev. D
  \textbf{57} 5480--5492 (1998).

\bibitem{RevModPhys.72.1149}
 C.~J. Hogan, Rev. Mod. Phys. \textbf{72} 1149--1161 (2000).

\bibitem{PhysRevD.67.043517}
 J.~J. Yoo and  R.~J. Scherrer, Phys. Rev. D \textbf{67} 043517 (2003).

\bibitem{Martin:1997ns}
 S.~P. Martin, A supersymmetry primer. fourth edition (2006)
  [\href{http://arXiv.org/abs/hep-ph/9709356v4}{{\tt hep-ph/9709356v4}}].

\bibitem{Hill:2002ap}
 C.~T. Hill and  E.~H. Simmons, Phys. Rept. \textbf{381} 235--402 (2003)
  [\href{http://arXiv.org/abs/hep-ph/0203079}{{\tt hep-ph/0203079}}]
  [Erratum-ibid. \textbf{390} 553-554 (2004)].

\bibitem{Chanowitz:2004gk}
 M.~S. Chanowitz, Czech. J. Phys. \textbf{55} B45--B58 (2005)
  [\href{http://arXiv.org/abs/hep-ph/0412203}{{\tt hep-ph/0412203}}].

\bibitem{Schmaltz:2005ky}
 M. Schmaltz and  D. Tucker-Smith, Ann. Rev. Nucl. Part. Sci. \textbf{55}
  229--270 (2005) [\href{http://arXiv.org/abs/hep-ph/0502182}{{\tt
  hep-ph/0502182}}].

\bibitem{Csaki:2005vy}
 C. Csaki,  J. Hubisz and  P. Meade, Electroweak symmetry breaking from extra
  dimensions, in J.~Terning, C.~E.~M. Wagner and D.~Zeppenfeld (Eds)  {\itshape
  Physics in $D \ge 4$: TASI 2004}  (World Scientific, Singapore, 2006), pp.
  703--776 [\href{http://arXiv.org/abs/{hep-ph/0510275}}{{\tt
  {hep-ph/0510275}}}].

\bibitem{Djouadi:2005gi}
 A. Djouadi, The anatomy of electro-weak symmetry breaking. I: The Higgs boson
  in the standard model. (2005)
  [\href{http://arXiv.org/abs/hep-ph/0503172}{{\tt hep-ph/0503172}}].

\bibitem{Dimopoulos:1979es}
 S. Dimopoulos and  L. Susskind, Nucl. Phys. \textbf{B155} 237--252 (1979).

\bibitem{Eichten:1979ah}
 E. Eichten and  K.~D. Lane, Phys. Lett. \textbf{B90} 125--130 (1980).

\bibitem{Djouadi:2005gj}
 A. Djouadi, The anatomy of electro-weak symmetry breaking. II: The Higgs
  bosons in the minimal supersymmetric model. (2005)
  [\href{http://arXiv.org/abs/hep-ph/0503173}{{\tt hep-ph/0503173}}].

\bibitem{Lee:1977eg}
 B.~W. Lee,  C. Quigg and  H.~B. Thacker, Phys. Rev. \textbf{D16} 1519 (1977).

\bibitem{Cornwall:1974km}
 J.~M. Cornwall,  D.~N. Levin and  G. Tiktopoulos, Phys. Rev. \textbf{D10} 1145
  (1974)  [Erratum-ibid.~\textbf{D11} 972 (1975)].

\bibitem{Chacko:2005pe}
 Z. Chacko,  H.~S. Goh and  R. Harnik, Phys. Rev. Lett. \textbf{96} 231802
  (2006) [\href{http://arXiv.org/abs/hep-ph/0506256}{{\tt hep-ph/0506256}}].

\bibitem{Amaldi:1991cn}
 U. Amaldi,  W. {de Boer} and  H. F\"{u}rstenau, Phys. Lett. \textbf{B260}
  447--455 (1991).

\bibitem{Spergel:2006hy}
 D.~N. Spergel {\itshape et~al.}, Wilkinson Microwave Anisotropy Probe (WMAP)
  three year results: Implications for cosmology. (2006)
  [\href{http://arXiv.org/abs/astro-ph/0603449}{{\tt astro-ph/0603449}}].

\bibitem{Bertone:2004pz}
 G. Bertone,  D. Hooper and  J. Silk, Phys. Rept. \textbf{405} 279--390 (2005)
  [\href{http://arXiv.org/abs/hep-ph/0404175}{{\tt hep-ph/0404175}}].

\bibitem{cdf}
{The Collider Detector at Fermilab}
  [\href{http://www-cdf.fnal.gov}{\texttt{www-cdf.fnal.gov}}].

\bibitem{d0}
{The D\O\ Experiment} [\href{http://www-d0.fnal.gov}{\texttt{www-d0.fnal.gov}}].

\bibitem{tevlum}
{Tevatron Luminosity Charts}
  [\href{http://www.fnal.gov/pub/now/tevlum.html}{\texttt{www.fnal.gov/pub/now%
/tevlum.html}}].

\bibitem{lhc}
{The Large Hadron Collider Project}
  [\href{http://lhc.web.cern.ch/lhc}{\texttt{lhc.web.cern.ch/lhc}}].

\bibitem{Gianotti:2005fm}
 F. Gianotti and  M.~L. Mangano, LHC physics: The first one-two year(s), in
  G.~Carlino and P.~Paolucci (Eds)  {\itshape Proceedings of the 2nd Italian
  Workshop on the Physics of Atlas and CMS}, Frascati physics series 38  (INFN,
  Frascati, 2005), pp. 3--26 [\href{http://arXiv.org/abs/hep-ph/0504221}{{\tt
  hep-ph/0504221}}].

\bibitem{atlas}
{The ATLAS Experiment}
  [\href{http://atlasexperiment.org}{\texttt{atlasexperiment.org}}].

\bibitem{cms}
{Compact Muon Solenoid} [\href{http://cms.cern.ch}{\texttt{cms.cern.ch}}].

\end{thebibliography}


\end{document}